# Micromachined Inclinometer Based on Fluid Convection

N. Crespy[1], J. Courteaud[1], P. Combette[1], P. Temple Boyer[2], A. Giani[1], A. Foucaran[1]

[1]IES: Institut Electronique du Sud
5 place Eugène Bataillon
34095 Montpellier, France

[2]LAAS: Laboratoire d'Analyse et d'Architecture des Systèmes
7 avenue du colonel Roche
31077 Toulouse, France

*Abstract*-This paper presents the experimental and theoretical study of an inclinometer based on heat exchange. The principle of the sensor is composed of a resistance heater placed between two detectors and suspended in a cavity filled with fluid. In this study, several fluids such as liquid or gaseous, were used. The temperature gradient and sensitivity as a function of fluid were studied by using the numerical resolution of equations of fluid dynamics with the Computational Fluid Dynamics (CFD) software Fluent V6.2. We showed that the sensitivity of the "liquid sensors" is greater than "gas sensors" one. The experimental measurements corroborate with numerical simulation. In addition, we have demonstrated that the sensitivity of the sensor is proportional to the Rayleigh number which is characteristic of the natural convection flow.

I. INTRODUCTION

The inclinometer is a special type of the accelerometer. It is gravity which acts on the sensor, as shown in Fig. 1. It is widely used in the field of military industries, robotics systems, seismic monitoring and particularly in automotive applications, such as chassis regulation and over-roll detection. The inclinometer presented in this paper resists to high accelerations (about 50000 g) because it have no proof mass and a small size because of its micromachined elaboration.

This study presents the experimental results and numerical simulations of a one-dimensional thermal inclinometer. The principle of the sensor is based on the heat exchange of a suspended resistance heater placed between two detectors. When the resistance is electrically powered, it creates a symmetrical temperature gradient within a micromachined silicon cavity filled with gas or liquid. The angle of inclination is deducted from the measure of the acceleration due to the gravity projected on the sensitive axis (Fig. 1).

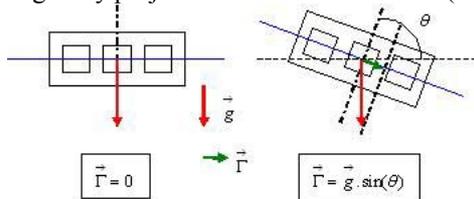

Fig. 1. The sensor principle.

II. THEORY

In the present study, we consider that the heat exchange in the fluid occurred principally by natural convection and weakly by radiation. The natural convection is governed by the buoyancy forces and this phenomenon between a surface and a fluid may be considered as a conduction problem in an moving environment. The equations to be implemented are those of fluid mechanics and those of conduction. However, in a flow in contact with a wall temperature $T_p$, there is a thin layer of viscous fluid which is considered as laminar [1]. This area is called the thermal boundary layer. Within this zone, the maximum temperature variation is defined by:

$$0.99(T_p - T_0) \quad (1)$$

With $T_p$ temperature of the wall and $T_0$ reference temperature.

We admit that there is no mixing of material in the direction perpendicular to the wall and the heat is preferably transmitted by conduction in the boundary layer.

Outside of this zone, the heat is transmitted by mixing of the fluid particles, causing rapid equalization of temperature. The heat flux $\phi$ across the boundary layer obeys the Fourier's law:

$$\phi = \frac{\lambda}{\delta_{th}}.S.(T_p - T_0) \quad (2)$$

With $\lambda$ thermal conductivity, $\delta_{th}$ thickness of the thermal boundary layer, S surface.

However, the problems related to thermal convection are complex, it is often impossible to know the thickness of the boundary layer. In natural convection, we use a modified form and empirical understanding of the law Fourier:

$$\phi = h.S.(T_p - T_0) \quad (3)$$

With h convection coefficient

The determination of the coefficient of convection is not easy because it is not constant and it depends on many parameters. Generally, it is expressed globally for the whole surface, and so it is an average value for the system. As

     

précised previously, it varies locally and depends on the nature of the fluid and the surrounding temperature (it increases with temperature), the fluid's velocity of circulation in the vicinity of the plate (it increases with speed), the orientation (vertical or horizontal) and dimensions of the surface. The experimental study of all of these properties is not achievable and therefore dimensional analysis is used to group the parameters that influence a convection phenomenon.

For the dimensional analysis, five dimensionless numbers are deduced which are the Reynolds number Re, the Nusselt number Nu, the Prandt number Pr, the Grashof number Gr and the Rayleigh number Ra.

The Reynolds number is defined by the ratio of inertial forces to viscous forces. It also determines the flow regime and it is expressed by the formula:

$$\text{Re} = \frac{\rho . u_m . l}{\mu} \qquad (4)$$

With $\rho$ density, $u_m$ average speed of fluid flow, $l$ characteristic dimension of the system and $\mu$ dynamic viscosity.

As the energy transfer by convection is very closely linked to the movement of the fluid, it is necessary to know the mechanism of fluid flow before considering that of the heat flow. One of the most important aspects of the hydrodynamics study is to establish whether the movement of the fluid is laminar or turbulent. The flow regime is laminar if Re <2000, it is "intermediate" when 2000 <Re <3000 and is turbulent when Re> 3000.

The Nusselt number is the ratio between the amounts of heat exchanged by convection to the amount of heat exchanged by conduction involving the coefficient of convection $h_c$ and it is expressed by the formula:

$$Nu = \frac{h_c . l}{\lambda} \qquad (5)$$

The Prandtl number characterizes the distribution of velocities relative to the temperature distribution. It is a characteristic thermal physics of fluid. It is expressed by:

$$\text{Pr} = \frac{\mu . C_p}{\lambda} \qquad (6)$$

With $C_p$ specific heat capacity

The Grashof number quantifies the ratio and the importance of the strength of viscosity, the inertial force and the Archimedes' thrust to determine the behavior of the system. It is expressed by the formula:

$$Gr = \frac{\rho^2 . g . \beta . \Delta T . l^3}{\mu^2} \qquad (7)$$

With g acceleration or gravity, $\beta$ coefficient of thermal expansion, $\Delta T$ temperature difference between the hot plate and the reference.

The Rayleigh number is used to characterize the natural convection flow in the same way as the Reynolds number in forced convection. It expresses the relationship between the Archimede force divided by the product of the viscous drag and the rate of heat diffusion. It is expressed by:

$$Ra = \text{Pr} . Gr = g . l^3 . \Delta T . \frac{\rho^2 . \beta . C_p}{\mu . \lambda} \qquad (8)$$

If Ra < 47000, the heat exchanges are governed by convection in laminar flow and/or by conduction through the fluid. If Ra > 47000, the convection in turbulent flow takes place in the fluid.

So, first of all, we have to check for each fluid that the exchanges are done in laminar flow. This requires Ra inferior to 47000, and this result is founded for all the considered fluids (table I).

TABLE I
RAYLEIGH NUMBER OF EACH FLUID

| Fluid | He | $N_2$ | $CO_2$ | SAE50 | Rhodorsil 47V100 | Ethylene Glycol |
|---|---|---|---|---|---|---|
| Ra | $3.10^{-7}$ | $2.10^{-5}$ | $7.10^{-5}$ | $3.10^{-2}$ | $9.10^{-2}$ | $8.10^{-1}$ |

For inclinometer filled with gas, a simple model based on the Grashof number (6) was first proposed by Leung [3] and then was verified in others works [4-5-6] considering that the sensitivity of the thermal inclinometer is directly proportional to this number.

However, on liquid inclinometers few papers present results but no study indicating a model has been published [7-8]. As a function of temperature, the laws of variations on the specific properties of liquids do not follow the same ones as those of gases. More over, they are different for each liquid. Indeed, the gas can be considered as ideal, and therefore we know the laws of variation with temperature, viscosity, conductivity coefficient of expansion and density. For liquids, the influence of temperature occurs more complicated because the previous characteristics vary according to more complex laws than gases and depending on the nature of the liquid.

In the next chapter, the experimental study conducted on the liquid inclinometer, will be exposed. We will demonstrate that the sensitivity (S) of a liquid inclinometer is proportional to the Rayleigh number. Indeed, the model based on Grashof number is not complete enough then liquids are used because the parameters attached to the velocities are not taken into account. Relating to relation (7) and if we remember that for gases, the Prandtl number is constant, it can explain that the behavior of "gas inclinometer" is modelized by Gr expression.

III.  EXPERIMENTATION AND SIMULATION

For our measurements, a volume of cavity equal to 5 mm$^3$ was used. The resistance heater is 70 μm wide and the detectors are 7 μm wide and are positioned at 200 μm away from the heating resistance.

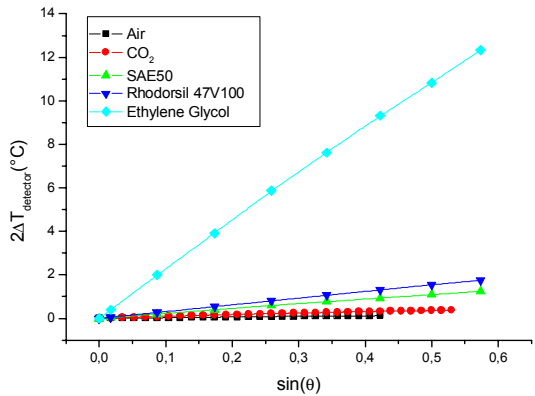

Fig. 2. Temperature Difference between 2 detectors based on sinus of the angle.

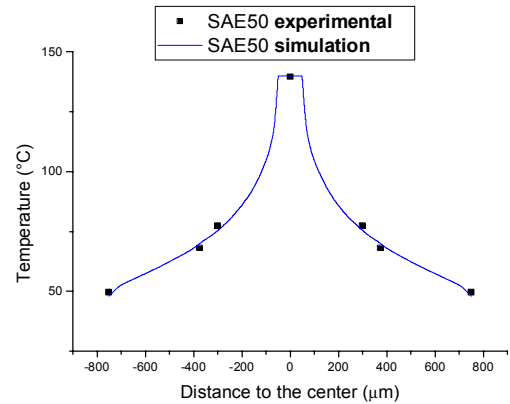

Fig. 4. Temperature gradient experimental and simulate for SAE50.

The temperature difference between the two detectors according to the sine of the angle was measured for air, $CO_2$, mineral oil SAE50 and Ethylene Glycol (Fig. 2). As expected, the sensitivity of the sensor depends on the nature of the fluid and the behavior is linear with an angle in the studied range.

The temperature gradient and sensitivity, shown in Fig. 3 and 4, were studied by using numerical resolution of equations of fluid dynamics with "Computational Fluid dynamics" (CFD) software Fluent 6.2. As we can see in the figures, the experimental results for mineral oil SAE50, are in good accordance with the simulation.

Then, the effect of temperature dependence versus thermal physics properties of fluids was discussed. In the previous chapter, we assume that the sensitivity (S) of liquid inclinometers is proportional to the Rayleigh number. We are going to demonstrate this here. If we take a sensor with the same rise in temperature for different liquids, then the parameters which vary, depending on the liquid, are reduced to:

$$S\ prop.\ Ra\ prop.\ \frac{\rho^2.\beta.C_p}{\mu.\lambda} = A \quad (9)$$

Table II summarizes the relative sensitivity of the sensor according to a number of different fluids. As we can see, the devices that present the greatest number A, i.e. liquids, have the greatest sensitivities.

TABLE II
SUMMARY OF INCLINOMETERS SENSITIVITY

| Fluid | He | $N_2$ | $CO_2$ | SAE50 | Rhodorsil 47V100 | Ethylene Glycol |
|---|---|---|---|---|---|---|
| $S/S_{air}$ | 0.02 | 1.01 | 3.04 | 7.18 | 9.55 | 71.43 |
| A | 516 | $35,5.10^3$ | $118,6.10^3$ | $57,6.10^6$ | $157,2.10^6$ | $1453,6.10^6$ |

The dominant parameter for liquid is viscosity. So the viscosity was plotted as a function of temperature and it was assimilated to a graph of the form $T^x$ (Fig. 5). The specific properties are derived from books and suppliers' specifications. Therefore, the parameter x was determined for each liquid.

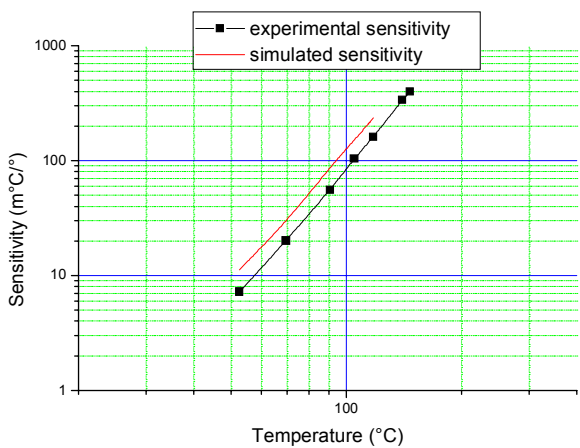

Fig. 3. Simulated sensitivity for oil SAE50 versus the heater temperature.

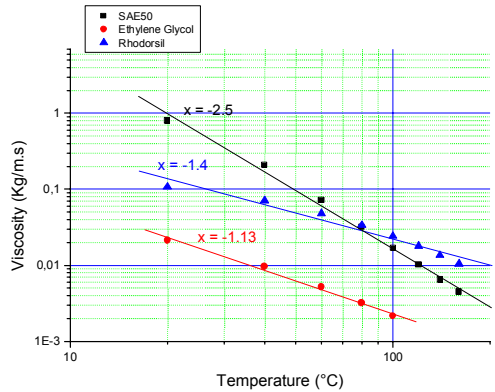

Fig. 5. Viscosity for different fluids as a function of temperature.

The sensor sensitivity was experimentally measured as a function of the ambient temperature for various liquids, keeping the same rise in temperature of 58 °C and the same average temperature of the heater (Fig. 6). The straight line corresponding to the previous results found for the viscosity, were plotted as a function of temperature on the same graph to verify that the sensitivity behavior is inversely proportional to viscosity (Fig. 6). As we can see, the slopes present an opposite value than those presented in graph of figure 5. So the sensitivity is inversely proportional to the viscosity and it is therefore proportional to the Rayleigh number.

We also verified that the sensitivity depends on the nature of the fluid. In the same conditions than the previous ones, the sensitivity was measured for different liquids. We have plotted the sensitivity for each liquid depending on their viscosity (Fig. 7). A slope equal to -1 was obtained, hence the sensitivity depends on the nature of the fluid and more particularly on the viscosity.

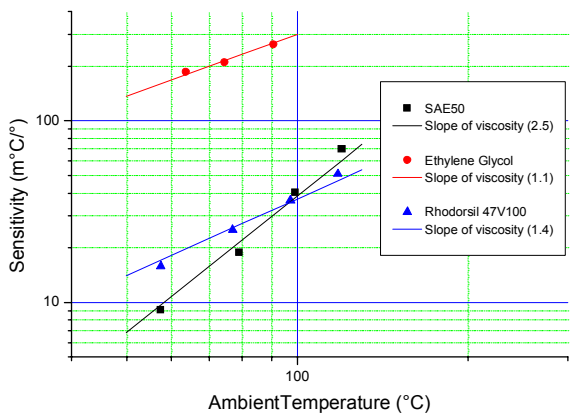

Fig. 6. Sensitivity for different fluids as a function of ambient temperature.

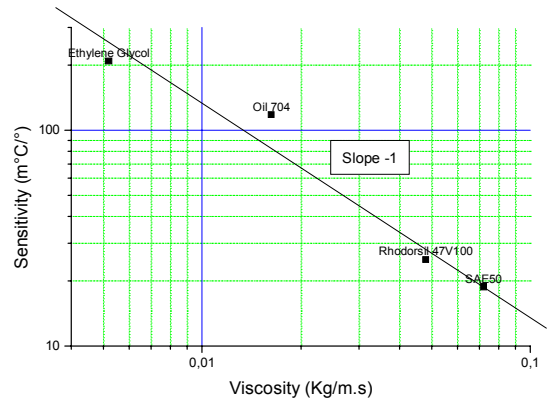

Fig. 7. Sensitivity as a function of viscosity.

In our demonstration, some differences can be seen between the model and experiment results. This is certainly due to the size of our sensor. Indeed, the width of the cavity is an important parameter since it depends on the size of the boundary layer. If we want to eliminate the edge effect to the operation of the sensor, the width of cavity should be larger than the boundary layer. The thermal boundary layer is the area where the heat transfer is the most important. For $CO_2$ gas with a rise in temperature of 250 °C on heating resistance, it has been calculated as 745 μm ± 10 μm. The cavity used was about 900 μm wide and is therefore well adapted to this fluid. The same calculation has been done for mineral oil SAE50 and for a temperature rise of the heater resistance equal to 140 °C. The thermal boundary layer is founded close to 3000 μm. So the sensor is not well optimized for this fluid and a larger will be used to improve the understanding of the model.

The response time of the sensor with different gases and different liquids was also measured. We applied a step excitation on the sensor and we measured the response time as it corresponds to a first order system (Fig. 8). As expected, the gas inclinometers have a response time lower than the liquid inclinometers. This is because the liquid is heavier and therefore its inertia is more consequent.

The table III represents the response time ($t_r$) of the sensors.

TABLE III
RESPONSE TIME OF THE SENSOR

| Fluid | He | $N_2$ | $CO_2$ | SAE50 | Rhodorsil 47V100 | Ethylene Glycol |
|---|---|---|---|---|---|---|
| $t_r$ (ms) | 0.47 | 2.3 | 5.3 | 480 | 500 | 510 |

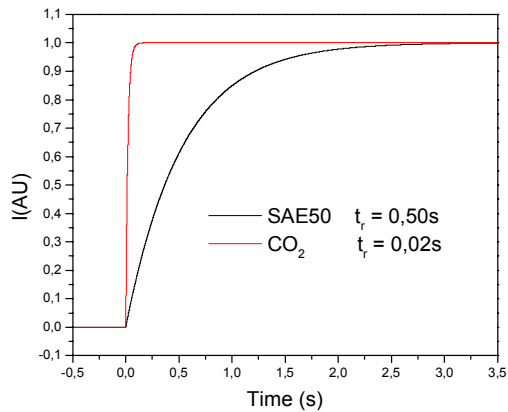

Fig. 8. Response time of one sensor.

## IV. CONCLUSION

So a comparative study of micro inclinometers filled with fluids of different sorts, has been presented. The sensitivity of the sensors has been simulated and measured for various gases and liquids. The behavior of the gas and liquid inclinometer can be modeling by the Rayleigh number. This is the first time that a model is reported on a study of inclinometer with a cavity filled with liquid. A good accordance was found between experimental results and the model. We were able to demonstrate an increase in the sensitivity of these devices compared to those filled with gas but their response time is shorter.